\documentclass[twocolumn,english,aps,prl,showpacs,superscriptaddress,amssymb,amsfonts]{revtex4}
\usepackage[T1]{fontenc}
\usepackage[latin9]{inputenc}
\usepackage{amsmath}
\usepackage{graphicx}
\usepackage{amssymb}
\usepackage{color}

\makeatletter
\newcommand{\be}{\begin{equation}}
\newcommand{\ee}{\end{equation}}
\newcommand{\bea}{\begin{eqnarray}}
\newcommand{\eea}{\end{eqnarray}}

\newcommand{\sz}{\sigma^{z}}
\newcommand{\sx}{\sigma^{x}}
\newcommand{\mz}{\mu^{z}}
\newcommand{\mx}{\mu^{x}}

\makeatother

\usepackage{babel}

\makeatother

\usepackage{babel}

\begin{document}

\title{Quantum Criticality in Topological Insulators and Superconductors: Emergence of Strongly Coupled Majoranas and  Supersymmetry}

\author{Tarun Grover}
\affiliation{ Department of Physics, University of California, Berkeley CA 94720}

\author{Ashvin Vishwanath}
\affiliation{ Department of Physics, University of California, Berkeley CA 94720}
\begin{abstract}
We study symmetry breaking quantum phase transitions in topological insulators and superconductors where the single electron gap remains open in the bulk. Specifically, we consider spontaneous breaking of the symmetry that protects the gapless boundary modes, so that in the ordered phase these modes are gapped. Here we determine the fate of the topological boundary modes right at the transition where they are coupled to the strongly fluctuating order parameter field. Using a combination of exact solutions 
and renormalization group techniques, we find that the surface fermionic modes either decouple from the bulk fluctuations, or flow to a strongly coupled fixed point which remains gapless. In addition, we study transitions where the critical fluctuations are confined only to the surface and find that in several cases the critical point is naturally supersymmetric. This allows a
 determination of critical exponents and points to an underlying connection between band topology and supersymmetry.
 Finally, we study the fate of gapless Majorana modes localized on point and line defects in topological superconductors at bulk criticality, which is analogous to a quantum impurity problem. Again, an interplay of topology and strong correlations causes these modes to remain gapless but in a strongly coupled state. Experimental candidates for realizing these phenomena are discussed.
\end{abstract}

\maketitle



\begin{figure}
\begin{centering}
\includegraphics[scale=0.4]{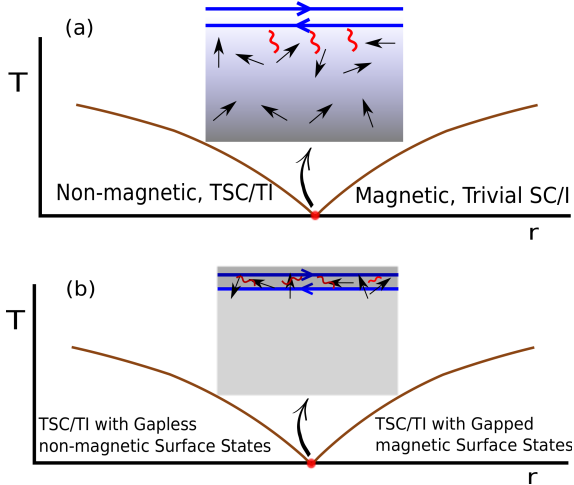}
\par\end{centering}
\caption{ The two class of problems studied in this paper that concern the fate of fermionic surface modes in TI/TSC at a critical point. $T$ denotes temperature while $r$ is the tuning parameter for
 the quantum phase transition. The counter-propagating fermionic surface modes (blue lines) interact with the critical fluctuations (headed vectors). The critical point may correspond to a bulk transition as depicted in (a)
or it could be a surface transition where the bulk is unaffected as shown in (b).} \label{fig:surfacetrans}
\end{figure}

The experimental verification of $\mathbb{Z}_2$ topological insulators in materials with strong spin-orbit coupling\cite{zhang2007, hseih2008, hseih2009, hseih2009_2, xia2009, chen2009, roushan2009} has lead
 to an explosion of interest in topological phenomena in solids. The theoretical description of these phases is now well understood, since the key properties such as the protected edge states are captured
 by models of non-interacting fermions\cite{TIreview, kane2005, zhang2006, fu2007, moore2007, roy2009, zhang2008, schnyder2008, kitaev2009, ryu2010}. Experimental realizations of these states in various bismuth based
 materials typically feature weak electron correlations, and are well described
 by conventional band theory.  However, recent work has begun to study the interplay of electron correlations and band topology, either by introducing correlated transition metal ions
such as Mn, Fe and Cu in bismuth based topological insulators \cite{SurfaceMagneticImpurities1, SurfaceMagneticImpurities2}, or by studying correlated materials with strong spin-orbit interactions
such as iridium based oxides and f-electron systems
\cite{half-heusler, iridates, osmates, topologicalKondoInsulators}.
Numerical studies of model correlated systems such as Kane-Mele-Hubbard Hamiltonian \cite{assaad2011, assaad2012} have also appeared. An important question for theory then is - are there
 qualitatively new phenomena that emerge from strong correlations, that are not captured by perturbing free electron topological states?

A novel feature of $\mathbb{Z}_2$ topological insulators, in contrast to integer quantum Hall states, is that they are well defined only in the presence of time reversal symmetry. A complete
 classification \cite{schnyder2008, kitaev2009} of free fermion topological phases reveals a superconducting analog of these states, which akin to topological insulators lack spin rotation symmetry,
 and which are also only well defined only in the presence of time reversal symmetry (class DIII). The well known B phase of superfluid He-3 is a realization of a topological phase in this
 class \cite{He3, volovikbook1, volovikbook2, roy2008, qi2008}, and possible solid state realizations have also been proposed \cite{fu2010,Leetriplet,RaghuAV}. In such symmetry protected topological phases, spontaneous
symmetry breaking provides a new route to exit the topological phase, without closing  the fermionic energy gap in the bulk. For example, electron correlations often lead to magnetic order,
 which breaks time reversal symmetry. A natural question concerns the evolution of the topological phase as magnetic order sets in. Therefore we consider quantum phase transitions from a paramagnetic topological phase, protected by time reversal symmetry, to a magnetically ordered phase, where
the topological surface states acquire a gap.  Although critical properties of the bulk are unaffected by the band topology, critical fluctuations near the surface will
interact with topological surface modes.  Note that these magnetic fluctuations have a significant impact on the protected surface modes when they condense, leading to a gap.
Do the surface states survive at the critical point, and if so are they fundamentally different from the free fermion states of the paramagnetic phase? Since critical  fluctuations
 are strongly interacting, particularly in low dimensions, answering this question typically  requires non-perturbative techniques. We use a combination of exact solutions,
renormalization group and mapping to models with enlarged symmetry, to study class DIII topological superconductors (DIII-TSC) in d=1,2,3 spatial dimensions and topological insulators in d=2,3. As usual, our results apply to a finite temperature `quantum critical' fan of the phase diagram, which is controlled by the zero temperature critical point\cite{Sachdev}.

Our main results are: (i) At the bulk critical point, the surface modes of the d=1 time reversal symmetric (DIII) TSC, which are a pair of Majorana zero modes, remain gapless but  couple strongly to the critical
fluctuations. This leads to edge fermion correlation functions that depart qualitatively from that of free Majorana fermions. We calculate their precise form using an exact solution.
 (ii) However, for DIII-TSCs in d=2,3 or TIs in d=3, the gapless surface modes decouple from the critical fluctuations at low energy, and are qualitatively unaffected at the critical
point. These results depend crucially on the fact that surface exponents of a transition are rather different from those in the bulk. (iii) We also consider the case when the surface
orders magnetically before the bulk. In particular we study the boundaries of DIII-TSC in d=2,3, when magnetic order sets in on crossing a quantum Ising critical point. These systems are found to display the remarkable feature of emergent supersymmetry at the critical point. The gapless critical modes of the
 Ising transition are superpartners of the Majorana surface states. In d=2 we argue that this is the tricritical Ising transition, accessed however by tuning a single parameter to attain
 surface criticality. Similarly, the d=3 system is argued to realize a supersymmetric critical point by tuning just a single parameter to attain surface criticality. Since this is one of the simplest settings to realize supersymmetry in a condensed matter system, we also discuss a number of striking physical consequences.
A similar phenomena occurs at the surface of TIs in d=3, at the critical point into a paired superconductor, which however requires tuning more parameters (doubled versions of this were
considered previously in different contexts \cite{sungsik,balents1998}). Finally (iv) we discuss the fate of topologically protected modes at the cores of point and line
defects, at the critical point. These open up a class of problems analogous to quantum impurity (or Kondo) problems, coupled to bosonic/fermionic bath
 \cite{wilson1975, andrei1983, hewson2000, sengupta2000, sachdev1999, vojta2007}, where the defect modes play the role of the impurity, coupled to bulk critical
fluctuations. However, these defects could also be one dimensional modes propagating along the defect line and therefore, apart from being of interest to the physics of TI/TSC, these problems
also serve to generalize the quantum impurity problem
to higher dimensional 'impurities', that are now one dimensional objects. For DIII-TSC we find  that the point (in d=2) and line defects
 (in d=3) remain gapless at the critical point but are essentially modified from the free Majorana form. The precise form of the strong coupling theory in d=2 is accessed using an $\epsilon (= 3-d)$ expansion.

We briefly review related prior work. Quantum phase transitions from a TI to a metal or trivial insulator, while preserving symmetries were studied in Ref.\cite{localization,localization1}. These transitions are
 rather different from the one described here, since they involve an intermediate state with gapless delocalized fermions in the bulk. In the problems we consider, the fermions always remain
 gapped while a bosonic symmetry breaking order parameter condenses. 
 Transitions where the symmetry protecting the topological phase is spontaneously broken have also been considered previously, but their effect on the surface modes has not been discussed. Purely surface magnetic transitions in TIs were discussed by Xu \cite{xu2010}, but the surface magnetic phase transitions in DIII-TSC, which turn out to have emergent supersymmetry as we show in this paper, were not considered. Surface superconducting transitions in 3dTIs, which also turn 
out to be supersymmetric when chemical potential is also tuned to obtain Dirac point, have also not been discussed earlier.
Quantum criticality in
 the Kane-Mele Hubbard model has  been discussed in Refs.\cite{xu2006, lehur2010,dunghailee2011,medhi2011, xu2012}  but again the impact on surface modes have so far been ignored. We now begin with
 a summary of the models and other preliminaries required for the remainder of the paper.
\bigskip

\section{Models and Main Results}

In this section, we provide a brief summary of the systems studied in this paper along with a synopsis of our main results. Topological insulators (TI) and topological superconductors (TSC) are recently discovered systems
 where the topology of
 underlying electronic band-structure leads to protected surface modes \cite{TIreview}. There are five different classes of TI/TSC in
every dimension where different classes differ from each other based on the underlying symmetry (time reversal, particle-hole etc.) \cite{TIreview, schnyder2008, ryu2010}.
Each class can itself contain a countably infinite number of distinct phases (`$\mathbb{Z}$ classification'), two phases (`$\mathbb{Z}_2$ classification') or only one distinct phase (`$\mathbb{Z}_1$ classification').
We focus on two classes: AII topological insulators (or simply topological insulators) and DIII topological superconductors, both of which require time-reversal symmetry. AII has the classification $\mathbb{Z}_1, \mathbb{Z}_2, \mathbb{Z}_2$ in
dimensions $d=1,2,3$ respectively while
DIII is classified as $\mathbb{Z}_2, \mathbb{Z}_2, \mathbb{Z}$ in $d=1,2,3$ respectively.

 We are primarily interested in the fate of boundary modes in these particular TI and TSCs at a magnetic and/or
superconducting symmetry breaking quantum phase transition, where the gap to the single-electron excitations in the bulk remains non-zero at the transition. Therefore, let us briefly discuss the nature of
boundary states  in these topological phases away from the phase transition.

First consider AII TIs. In $d=1$, the classification is $\mathbb{Z}_1$ and hence there
are no protected edge modes. In $d=2$, the canonical example of an AII TI is a quantum spin-Hall insulator and a generic edge is described by an interacting helical Luttinger liquid \cite{xu2006,wu2006}.
In $d=3$, the AII class corresponds to $\mathbb{Z}_2$ topological insulators whose surface modes are given, in their simplest form, by a single species of non-interacting Dirac fermion \cite{fu2007, moore2007, roy2009}. Weak interactions are irrelevant for these surface modes.

In contrast, the DIII TSC have non-trivial edge modes in all three dimensions. In $d=1$, these edge modes correspond to a pair of Majorana modes localized at the two ends of the system, that are related to one another by time reversal symmetry\cite{qi2008}. In $d=2$, they
correspond to a non-interacting helical Majorana liquid.  Finally, in $d=3$, again one obtains two-dimensional non-interacting Majorana modes. The integer classification in $d=3$ is related to the minimum number of such Majorana surface modes required by bulk topology. Here we only consider the case with a single surface mode. In all three cases, the edge modes remain gapless and non-interacting for weak interactions.

Strong interactions can lead to two very different kinds of symmetry breaking phase transitions in these topological phases. In the first kind, the transition occurs throughout the system. In the second kind
only the surface undergoes phase transition while the bulk remains gapped.
We find that these two kinds  of transitions comparatively have very different influence on the fermionic edge modes. In the former case, the nature of critical fluctuations near the boundary
of the system is dictated by the so-called `boundary critical phenomena' \cite{binder1983, diehl1986}. The primary implication of boundary critical phenomena is that
the universal critical exponents associated with various quantities such as specific heat are rather different for the boundary fluctuations when compared with the bulk. This fact has important consequences for the fermionic
 edge/surface modes in TI/TSC at criticality. We find that in both
two and three dimensions, owing to the non-trivial boundary critical exponents, the fermionic modes at the boundary completely decouple from the critical fluctuations for weak coupling. Therefore one obtains
 non-interacting gapless fermionic
 boundary modes even at the criticality. Only in the case of
$d=1$ DIII class TSC does the bulk fluctuations modify the behavior of fermionic edge modes. In this case, we find that even though the Majorana edge modes $\chi_1, \chi_2$ couple strongly with
the critical fluctuations, they still remain gapless and their density of states $N(E)$ changes from being a delta
function $N(E)\sim \delta(E)$ away from the criticality, to $N(E)\sim {\rm constant}$ at the critical point. Since the electron creation operator is related to Majoranas through $c^{\dagger} = \chi_1+ i\chi_2$,
the knowledge of $N(E)$ allows one to calculate the local density of states of the electrons $N_e(E)$ at
 the end of the chain, which is precisely what would be measured by scanning tunneling microscopy (STM):

\be
N_e(E) = \textrm{Im} \int\, dt\, \,\,\langle c(r=0,t=0) c^{\dagger}(r=0,t)\rangle e^{i(E+i0)t}   \nonumber
\ee

The differential conductance $dI/dV$ in STM is given by $dI/dV \propto N_e(E=eV)$ where $V$ is the voltage difference between the tip and the sample and $I$ is the current.

The critical exponents associated with magnetic correlations $C(r,t) = \langle S(r,t)\,S(0,0) \rangle$ near the boundary can in principle be accessed by Nuclear Magnetic Resonance (NMR) from the relaxation rate $1/T_1$, which measures the
imaginary part of spin-susceptibility $\chi$ at temperature $T$:

\be
\frac{1}{T_1 T} = \textrm{Limit}(\omega \rightarrow 0) \,\,\, \int  d^dk \,\,\frac{\textrm{Im} \chi(k,\omega)}{\omega} \nonumber
\ee

$\textrm{Im} \,\chi$ is proportional to the fourier transform of the correlation function $C(r,t)$.

For the second kind of phase transitions, where only the surface undergoes phase transition, the fermionic modes live in the same dimension as the critical fluctuations and the `boundary critical phenomena' is not relevant.
Correspondingly, we find that the boundary fermionic modes are strongly coupled to the critical fluctuations at the phase transition in almost all cases. Perhaps most interestingly,
we find that an Ising magnetic phase transition in a DIII TSC has an emergent \textit{supersymmetry} at criticality at long wavelengths  without any additional fine-tuning in both two and
three dimensions! We recall that the conventional symmetries of a quantum mechanical system, such as a rotation or a translation,
 transform a boson into boson and a fermion into fermion, the generator of the symmetry thereby being a boson. The hallmark of supersymmetry is the invariance under rotation matrices that are themselves \textit{fermionic} in nature.
Therefore, supersymmetry rotates a fermionic operator into a bosonic one and vice-versa. This has important physical consequences for our system, namely, DIII class TSC at criticality. For example, it results in the
equality of anomalous dimension corresponding to the  magnetic order parameter with that of the Majorana fermion, precisely because the bosonic order parameter can be rotated into a fermionic Majorana via
supersymmetry. We also find emergent supersymmetry at a superconducting phase transition on the surface of a three-dimensional AII TI with chemical potential tuned so as to obtain a Dirac fermion coupled to critical superconducting fluctuations.


DIII TSCs also host gapless Majorana modes at certain point and line defects in $d=2$ and $d=3$ respectively \cite{qi2008,teo2010}. The Hamiltonian for these modes in the absence of any interactions is identical to that of the edge modes in $d=1$ and $d=2$ DIII TSC. We study the fate of these Majorana modes at bulk criticality and find that the point defect exhibits anomalous exponents, and is hence very different from a free Majorana mode. The line defect behaves like a `marginal' Majorana liquid at criticality, with logarithmic corrections to free Majorana correlation functions.

\section{Surface Fermionic Modes Coupled to Bulk Criticality}

\underline{(i) Majorana modes in one-dimensional TSC:}

\begin{figure}
\begin{centering}
\includegraphics[scale=0.7]{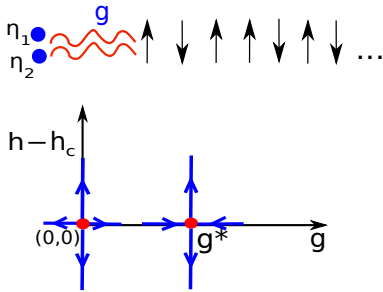}
\par\end{centering}
\caption{(above) Illustration of Majorana boundary modes $\eta_1, \eta_2$ in a one dimensional time-reversal invariant TSC coupled to critical Ising fluctuations with interaction strength $g$.
(below) Renormalization group flow for the same problem. $h$ is the transverse field for the Ising magnet. The $g=0$ fixed point is unstable to $g^{*} = 1$ fixed point where Majorana modes
are strongly coupled to the critical fluctuations and have correlation function $\langle \eta_\alpha(0) \eta_\alpha(\tau)\rangle \sim 1/\tau$ for $\alpha = 1,2$.} \label{fig:rg1dmajorana}
\end{figure}

 We begin by considering the simplest example in this class of problems: a one-dimensional TSC in class DIII at the criticality between TSC and an Ising magnetically ordered ordinary superconductor.
For simplicity we consider a semi-infinite wire of a TSC and consider the case of a finite chain in the Supplementary Information, Sec.A\ref{sec:suppl}.
The Hamiltonian of the system may be written as:

\bea
H & = & H_{Ising} + H_{Ising-Majorana} \nonumber \\
 & = & -\sum_{i=1}^{\infty} \left( \sigma_i^z \sigma_{i+1}^z + h \sigma_i^x \right) - i g \sigma_1^z \chi_1 \chi_2 \label{eq:Hdefect1d}
\eea

where $\chi_1$ and $\chi_2$ are the two Majorana modes located at the end of the wire. The fermionic degrees of the freedom are gapped in the bulk on either side of the phase
transition and do not play any role in the following discussion. Therefore we have not included them in the Hamiltonian $H$ (this holds for all phenomena considered in this paper).
When $h>1$, the Ising order parameter $\left< \sigma^z \right> = 0$ and the coupling $g$ between the Majorana modes
 and the Ising field $\sigma_1^z$ is  irrelevant, that is,
there are (free) Majorana modes at the end of the wire, as they should, in a TSC. On the other hand, when $h < 1$, $\left< \sigma^z \right> \neq 0 $, which provides a mass to the edge Majorana modes.
We are interested in the fate of the
Majorana modes right at the criticality ($h=1$). Using a classic result from the theory of boundary conformal field theory (BCFT) \cite{cardy1980, yellow}, the scaling dimension for the
boundary order spin $\sz_1$ is 1/2. Therefore the interaction term  $g i \int dt\, \chi_1(\tau) \chi_2(\tau) \sz(x=0, \tau)$ between the Majorana spins and the spin-chain is relevant
at the decoupled fixed point ($g=0$) in the renormalization group (RG) sense and hence it is an unstable fixed point. To proceed, let us define $\sz_0 = i \chi_1 \chi_2$.  Further, let us set $g=1$. We will soon consider the deviation of $g$
from unity.  Thus, at criticality, $H$ in Eqn.\ref{eq:Hdefect1d} may be rewritten as:

\be
H = -\sum_{i=1}^{\infty} \sigma^{x}_i - \sum_{i=0}^{\infty} \sigma^{z}_i \sigma^{z}_{i+1}
\ee

$H$ as written above corresponds to transverse field Ising model \textit{without} a transverse field for the first spin $\sigma_0$. Following \cite{masaki}, let us perform the following duality transformation:

\bea
\sigma^{x}_0 & = & \mu^{z}_0 \nonumber \\
\sigma^{x}_i & = & \mu^{z}_{i-1} \mu^{z}_{i} \hspace{0.5cm} \forall\,\, i \geq 1 \nonumber \\
\sz_i \sz_{i+1} & = & \mx_i \hspace{0.5cm} \forall\,\, i \geq 0 \nonumber
\eea

Hamiltonian $H$ in terms of the new variables $\mu$ is given by the standard transverse field Ising model:

\be
H = -\sum_{i=0}^{\infty} \left( \mx_i + \mz_i \mz_{i+1} \right) \nonumber
\ee

This form allows one to use the results from boundary conformal field theory to calculate the unequal time correlation function for Majorana fermions $\chi_1, \chi_2$. In particular, the spin-operator
$\sx_0$ may be defined as $\sx_0 = i \chi_3 \chi_1$ where $\chi_3$ is a Majorana operator without any dynamics. This definition ensures that $\sx_0$ anticommutes with $\sz_0$. Therefore, the time-correlation
functions of $\chi_1$ are same as that of operator $\sx_0 (= \mz_0)$:

\be
\langle \chi_1(0) \chi_1(\tau) \rangle = \langle \mz_0(0) \mz_0(\tau)\rangle   \sim 1/\tau \nonumber
\ee

since the scaling dimension of the boundary spin $\mz_0$ is 1/2 in Ising BCFT. Due to the symmetry, the correlation functions of $\chi_2$ are identical to that of $\chi_1$. Let us consider deviation of $g$ from unity: the term
$(g-1) i \chi_1 \chi_2 \sz_0 = (g-1) \mu_0^x$ turns out to be irrelevant because the boundary scaling dimension for the transverse field is $2$ \cite{yellow}. Therefore, $g=1$ is a stable
fixed point. We conclude that the interaction with the critical modes dramatically changes the scaling dimension of the Majorana modes from 0 to 1/2 and therefore they behave like a
zero-dimensional non-Fermi liquid. The density of states at energy $E$ changes from being a
delta function $\delta(E)$ in the non-interacting limit to a non-zero constant at the stable fixed point which can be probed using STM. The feedback of the Majorana fermions
on the Ising chain is also rather dramatic. When $g=0$, the edge correlations $\langle \sz_1(0) \sz_1(\tau) \rangle \sim 1/\tau$, as discussed above. Using the above duality, at $g=1$, that is, at the stable fixed point,
the operator $\sz_1 = \sz_0 \mx_0$. Since, $\sz_0$ commutes with the Hamiltonian and therefore does not have any dynamics, $\langle \sz_1(0) \sz_1(\tau) \rangle = \langle \mx_0(0) \mx_0(\tau) \rangle \sim 1/\tau^4$.
This follows from the fact that the scaling dimension of the energy operator at the boundary $(= \mx_0)$ is 2 \cite{yellow}.
This exponent can be measured from the imaginary part of the spin-susceptibility.
The renormalization group flow diagram is shown in Fig.\ref{fig:rg1dmajorana}

%
%
\underline{(ii)Majorana modes in two/three-dimensional TSC:}

\begin{figure}
\begin{centering}
\includegraphics[scale=0.5]{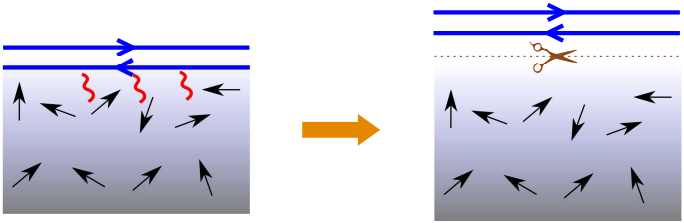}
\par\end{centering}
\caption{ The coupling to bulk critical fluctuations is irrelevant for two and three dimensional DIII TSCs and three dimensional AII TIs. This means that at low energies,
the boundary fermionic modes completely decouple from the bulk critical fluctuations.} \label{fig:decouple}
\end{figure}

Let us study analogous problems in higher dimensions. As a first example, consider the magnetic phase transition out of a two dimensional TSC in class DIII. The unperturbed edge states
consist of helical Majorana modes:

\be
H = \int dx\,\,i v_F \left( \chi_R \partial_x \chi_R - \chi_L \partial_x \chi_L \right) \label{eq:helicalmajorana1D}
\ee

At the critical point, these edge states are coupled to the critical magnetic fluctuations $\phi$ via a term $\Delta S$ in the action

\be
\Delta S = i g \int dx_{\parallel} d\tau \,\chi_L(\tau,x_\parallel) \chi_R(\tau,x_\parallel) \phi(\tau,x_\perp = 0, x_\parallel) \label{eq:coupling2d}
\ee
 where $x_\parallel (x_\perp)$ denote
spatial coordinates parallel (perpendicular) to the boundary of the system. To determine the relevance/irrelevance of this term at the $g=0$ fixed point,
similar to the case of Majorana mode above, it is
crucial to use the boundary critical exponents for the magnetic fluctuations $\phi$. The relevant exponents are known to a very good accuracy numerically \cite{night2005}. In
particular, the edge correlation functions $\langle \phi(x_\perp = 0, x_\parallel =0) \phi(x_\perp = 0, x_\parallel = L) \rangle $ decay approximately as \cite{night2005}

\be
\langle \phi(x_\perp = 0, x_\parallel =0) \phi(x_\perp = 0, x_\parallel = L) \rangle \sim 1/L^{2.52}
\ee

The non-interacting Majorana fermions in two space-time dimensions have a scaling dimension of 1/2. Therefore, in renormalization group sense, the coupling $g$ at length scale $L$ scales as
$g(L) \sim 1/L^{0.26}$  making it irrelevant at the $g=0$ fixed point. Therefore, for weak $g$, one obtains well defined non-interacting Majorana modes even at the critical point between the TSC and a magnetically ordered
trivial superconductor. Note that if one naively uses the bulk scaling dimension for $\phi$, which is close to 1/2, one would instead reach the erroneous conclusion that the coupling $g$ is relevant. This again underscores
the importance of boundary critical phenomena for the fate of topological surface modes at the bulk ordering transition.

Next, consider the same problem for a three dimensional DIII TSC. The boundary fermionic modes in this case are described by the Hamiltonian

\bea
H & = & \int d^2x \,\,i v_F \,\chi^T \left(\sigma_z \partial_x - \sigma_x \partial_y \right)\chi \label{eq:helicalmajorana2D}
\eea

where $\chi = [\chi_1 \,\, \chi_2]^T$ is a two-component Majorana spinor. The coupling between these surface modes and the critical fluctuations is given by an expression similar to
Eq.\ref{eq:coupling2d}:

\be
\Delta S = i g \int d^2x_{\parallel} d\tau \,\chi_L(\tau,x_\parallel) \chi_R(\tau,x_\parallel) \phi(\tau,x_\perp = 0, \vec{x}_\parallel) \label{eq:coupling2d}
\ee

where $\vec{x}_\parallel$ is now two-dimensional. In this case, the space-time dimension for the bulk is four,
which is the upper critical dimension for the critical fluctuations and therefore the boundary critical correlations exhibit mean-field behavior \cite{binder1983, diehl1986}:
$\langle \phi(x_\perp = 0, x_\parallel =0) \phi(x_\perp = 0, x_\parallel = L) \rangle \sim 1/L^4$. The scaling dimension of Majorana fermions in three space-time dimensions being unity,
this implies that $g$ is again irrelevant. Thus we find that the surface Majorana modes remain well-defined for small values of coupling to the critical modes for a
three dimensional superconductor as well.

\underline{(iii) Dirac modes in three dimensional TI:}

The analysis of three dimensional AII TIs where the chemical potential has been tuned to obtain Dirac fermions at its surface proceeds in the exact same manner as that for
three dimensional DIII TSCs and one finds that Dirac fermions remain
 well-defined at the criticality. Note that if the critical fluctuations were instead confined to the boundary (that is, a surface phase transition), the result would be entirely different.
In particular, for a surface transition the coupling to critical modes will be relevant and the Dirac fermions will not be free at the transition \cite{xu2010}.
For a generic chemical potential one obtains a helical Fermi liquid. Ref.\cite{xu2010} studied the problem of helical Fermi surface at a surface phase
transition and it was found that the coupling between helical Fermi liquid and critical fluctuations is irrelevant.
For bulk critical fluctuations, the problem of interest in this section, the Landau damping term has the same form as in
Ref.\cite{xu2010} $\sim \int d\omega d^2q_{\parallel} dq_\perp |\phi(q_\parallel, q_\perp, \omega)|^2 |\omega|q_\parallel$ and the coupling to critical fluctuations is even more strongly irrelevant
owing to the boundary critical exponents cited above.
 Therefore, the helical Fermi liquid is unaffected
by bulk critical fluctuations for weak coupling. The case of two dimensional TIs and the Kane-Mele Hubbard model requires a more careful analysis due to the Luttinger liquid nature of the edge   states and the results will be reported elsewhere.

\section{Surface Fermionic Modes Coupled to Surface Criticality}

In the previous section we studied the fate of fermionic boundary modes at bulk symmetry breaking quantum phase transitions. However, it is also conceivable that in certain situations the surface undergoes transition before the bulk does. In this section, we study the fate of surface modes at such surface phase transitions.

\underline{(i) Emergent supersymmetry in three-dimensional TSC:}

Let us consider a surface magnetic phase transition in a three dimensional DIII TSc. The unperturbed surface states are described by the Hamiltonian in Eqn. \ref{eq:helicalmajorana2D}.
At the transition, these surface states are coupled to an Ising magnetic order parameter. Note that one cannot couple the system to an XY or O(3) magnetic order parameter since the discrete time-reversal
is the only symmetry that the surface states possess.
A scenario such as this can arise in several ways. One can imagine a situation where the material has two bands, one of them contributes itinerant electrons which form a topological superconductor while the other
band consists of localized moments that provide the Ising magnetic variable. Alternatively, one can also dope the system with magnetic impurities in a uniform manner so that the disorder effects can be ignored.
What is the fate of the boundary states at such a surface phase transition
if the bulk still remains gapped? The low-energy theory of the Majorana modes coupled to
Ising critical fluctuations is:

\bea
S & = & \int d\tau\, dx\, dy \,\,\left[{\tilde{\chi}}^T \left( \sigma_y \partial_\tau + \sigma_x \partial_x + \sigma_z \partial_y \right) \chi + i g \,\phi\,\, \tilde{\chi}^T \chi + \right. \nonumber \\
& & \left. \frac{1}{2}{(\partial_\mu \phi)^2} + \frac{r}{2}\phi^2 + u \phi^4\right] \label{eq:susy3dtsc}
\eea

where $\tilde{\chi}^T = \left[ \chi_1 \,\,\, \chi_2 \right] i \sigma_y$. If $g$ were zero, the Ising order parameter will undergo a phase transition such that the critical point is described by
the conventional Wilson-Fisher fixed point \cite{wilson1972}. However, as shown in Ref.\cite{sonoda, thomas2005}, in the presence of the coupling $g$, the Wilson-Fisher fixed point is
unstable to a new conformally invariant fixed point that has an emergent $\mathcal{N} = 1$ supersymmetry. As mentioned earlier that supersymmetry allows one to rotate fermions into bosons
(and vice-versa) and the label $\mathcal{N}$ denotes the independent number of such fermionic rotation operators. Supersymmetry requires $u=g^2/8$ and one would have naively thought that satisfying this
relation requires fine tuning and hence the supersymmetric fixed point would be multi-critical. However, in this case, the supersymmetry is \textit{emergent} \cite{sonoda, thomas2005}, by which we mean that
 this relation is satisfied automatically at the fixed point and the only tuning parameter is the boson mass $r$, as in a conventional phase transition. Let us briefly consider the supersymmetric transformations
that keep the above action invariant \cite{wessbagger}:

\bea
\delta \phi & = & \tilde{\epsilon}^T \chi \nonumber \\
\delta \chi & = & ( \frac{\gamma_\mu  \partial_\mu \phi}{2} + i\frac{g}{4} \phi^2)\epsilon \nonumber
\eea

where $\gamma_\mu = \{\sigma_y, \sigma_x, \sigma_z\}$ and $\epsilon = [\epsilon_1\,\,\epsilon_2]^T$ is a two-component Majorana variable. The emergence of this `fermionic rotational invariance' implies that at the critical point the correlation function of the
Majorana fermions $\langle \tilde{\chi}^{T}(\omega, \vec{p}) \chi(-\omega, -\vec{p}) \rangle \sim |\omega^2 + \vec{p}^2|^{\eta/2} (\sigma_y \omega + \sigma_x p_x + \sigma_z p_y)^{-1}$
and the order parameter $\langle \phi(\omega, \vec{p}) \phi(-\omega, -\vec{p}) \sim (p^2 + \omega^2)^{\eta/2-1}$ have exactly the same scaling exponent $\eta$ \cite{sonoda}. Experimentally, this can be verified
by extracting the fermionic anomalous exponent using STM or ARPES while that corresponding to the order parameter can be extracted from the neutron scattering.

At finite temperature, the supersymmetry is spontaneously broken \cite{das1978, grisaru1981, buchholz1997} and as a result, one obtains a collective fermionic excitation, `phonino', an analog of
Goldstone boson, as first pointed out by Lebedev and
Smilga \cite{lebedev}.
Since the physics in quantum critical regime is controlled by the critical fixed point, the contribution of phonino to various thermodynamic and transport properties should be visible in this regime. We leave
the details of this interesting subject to future work.


\underline{(ii) Emergent supersymmetry in two-dimensional TSC:}

Let us consider a one lower dimensional analog of the problem studied above viz. boundary of a two dimensional
DIII TSc (Eqn.\ref{eq:helicalmajorana1D}) coupled to critical Ising magnetic fluctuations:

\bea
S & = & \int d\tau\,dx \,\, \left[ \left( \chi_R \partial_\tau \chi_R + \chi_L \partial_\tau \chi_L + i \chi_R \partial_x \chi_R - i \chi_L \partial_x \chi_L \right) + \right. \nonumber \\
& & \left. ( \frac{1}{2}(\partial_\mu \phi)^2 + \frac{r}{2} \phi^2 + \frac{u}{4} \phi^4) +  2 i g \chi_L \chi_R \,\phi \right] \label{eq:tricrit1}
\eea

In the absence of any coupling to the magnetic fluctuations, the edge theory of the Majorana modes is identical to that of a $1+1$-d critical Ising model self-tuned to criticality. Therefore when $g=0$
and the $\phi$ field is at Ising criticality, the above action describes two decoupled critical Ising models. The scaling dimension $\Delta_g$ of $g$ equals $\Delta_g = 1 + 1/4 = 5/4 < 2$, and therefore it is
a relevant perturbation at this decoupled fixed point. What is the fate of the coupled theory? Below we present several arguments that it flows to the tricritical Ising fixed point with conformal charge $c=7/10$.

Let us begin with a simple mean field analysis and denote the Ising variable corresponding to the bosonized Majorana modes as $\phi_1$ and define $\phi_2 \equiv \phi$.
At the mean field level, the coupled system may be described by the following Landau free energy:

\be
F = u_1 \phi^4_1 + v_1 \phi^6_1 + r_2 \phi^2_2 + u_2 \phi^4_2 + v_2 \phi^6_2 + g \phi_1^2 \phi_2 \label{eq:tricrit2}
\ee

$\phi^2_1$ term in $F$ is not allowed since the Majorana modes are self-tuned to criticality, being the boundary of a TSC. Furthermore, note that the $g$ term breaks the symmetry from $Z_2 \times Z_2$ to $Z_2$
as in Eqn.\ref{eq:tricrit1} and describes an energy-order parameter coupling. Assuming $\phi_2$ to be small near the critical point, one can eliminate
it in favor of $\phi_1 \Rightarrow \phi_2 \approx -\frac{\phi^2_1}{2 r_2}$. This leads to the following free energy (upto $O(\phi^6_1)$) for the variable $\phi_1$:
\be
F = (u_1 - \frac{1}{2r_2}) \phi^4_1 + v_1 \phi^6_1
\ee
When $r_2 \gg 1$, the minimum of $F$ lies at $\phi_1 = 0$ with $F$ vanishing as quartic power of $\phi_1$. This implies that the Majorana fermions are at Ising criticality while the field $\phi$ is gapped
consistent with the action in Eq.\ref{eq:tricrit1}. At $r_2 = 1/2u_1$, the potential for $\phi_1$ becomes $\sim \phi^6_1$ which corresponds to Ising model at \textit{tricriticality}.
For $r_2 < 1/2u_1$, the minima of $F$ lie at $\phi_1 \neq 0$ which corresponds to mass acquisition of Majorana due to condensation of $\phi$. Therefore, we conclude that within the mean-field theory, \textit{the
phase transition from a non-magnetic one-dimensional TSC to a magnetic superconductor is described by a tricritical Ising model!}

There are several consistency checks that support this conclusion.
First, Zamolodchikov's $c$-theorem \cite{zamol1986} implies that the central charge of
the system in Eq.\ref{eq:tricrit1} at the magnetic phase transition must be less than that of two decoupled Ising models ($=1/2+1/2 =1$). Indeed,
the central charge for the tricritical Ising model is $c = 7/10$, consistent with this constraint. Furthermore, if the magnetic phase transition is indeed in the tricrital Ising universality, owing to the $c$-theorem,
a relevant perturbation can either take it to a gapped fixed point ($c =0$) or Ising critical point ($c = 1/2$), since those are the only two allowed values of central charge for unitary theories that are less than 7/10.
This property is indeed shared by the action in Eq.\ref{eq:tricrit1}: giving a mass to $\phi$ takes the system to Ising critical point while condensing $\phi$ takes it to a gapped phase.
Interestingly, the tricritical Ising has emergent $\mathcal{N}=1$ supersymmetry \cite{friedan1, friedan2}, consistent with the matter content of the action in Eqn.\ref{eq:tricrit1}
which contains one real fermion and one real boson. The Majorana fermion $\chi$ and the scalar $\phi_2$ are superpartners of each other with the same anomalous dimension of 1/5 \cite{yellow}.
Like in other cases, the experimental predictions such as the scaling dimension of Majorana fermion and the Ising order parameter can in principle be accessed via single particle tunneling and NMR.

Interestingly, the flow from the supersymmetric Tricritical Ising model to
the Ising model  can be thought of as spontaneous breaking of the
supersymmetry.
Within this picture, the free Majorana fermion at the Ising critical point
can be thought of as the `goldstino' associated with the spontaneous
breaking of the supersymmetry.
This picture also generalizes to the higher dimensional supersymmetric
models that we study. Since the presence of free fermionic boundary modes
in the topological phase is a consequence of topological band structure,
which is well-defined even at the surface phase transition, this
observation hints
at an interesting interplay between band topology and emergence of
supersymmetry at the surface quantum critical points.

\underline{(iii) Emergent supersymmetry in three-dimensional TI:}

Next, we study surface phase transitions in a three dimensional AII TI. In the absence of interactions, the
Hamiltonian for a surface with normal $\vec{n} = \vec{z}$ is given by:

\be
H = \sum_{k_x, k_y} c^{\dagger} \left(k_x \sigma_y - k_y \sigma_x - \mu \right) c  \label{eq:helicalTI3D}
\ee

where $\mu$ is chemical potential. Let us consider fine-tuning $\mu$ to zero so as to obtain fermions with massless Dirac dispersion at the surface and study
the instability of these surface modes to an s-wave superconductor. Such a transition can potentially be driven by intrinsic interactions and may also be realized by pattering the surface with a Josephson junction array.
We further restrict ourselves to a case where the superconducting fluctuations are particle-hole symmetric and therefore, in the absence of coupling to the surface fermions, would exhibit a quantum
phase transition with dynamic critical exponent $z=1$ (this may require additional fine-tuning). Thereby the effective action for the coupled system near the phase transition may be written as

\bea
S & = & \int d^3x \,\,\left[\overline{\psi} \gamma_\mu \partial_\mu \psi + g\,\left(\phi\,\, \psi^T \epsilon \psi  + c.c.\right) \right. \nonumber \\
& & + \left. |\partial_\mu \phi|^2 + r|\phi|^2 + u |\phi|^4\right] \label{eq:susy3dTI}
\eea

where $\phi$ is the superconducting order parameter, $\psi^T = \left[c_{\uparrow} \,\,\,c_{\downarrow}\right]$,\, $\{\gamma_0,\gamma_1,\gamma_2\} = \{\sigma_z, \sigma_x,\sigma_y \}$ and $\epsilon$ is
 two-dimensional antisymmetric tensor. Akin to the case of DIII TSC coupled to Ising order parameter, the above action is also known to flow to a supersymmetric fixed point \cite{sungsik, thomas2005}
where the fixed point value of the couplings satisfy $g^2 = u$. In this case, the supersymmetry corresponds
 to a $\mathcal{N} = 2$ Wess-Zumino model which implies that there exist two fermionic generators that implement supersymmetry. Supersymmetry allows one to calculate the exact anomalous dimensions at this fixed point.
 In particular, $\eta_\phi$ and $\eta_\psi$ both equal $1/3$ \cite{aharony}.
To further explore the experimentally testable consequences of supersymmetry, we note that like any other symmetry, supersymmetry is associated with conserved currents (`supercurrents'), which in this case are given by \cite{wessbagger}:

\bea
J_\mu & = & (\gamma_\rho \partial_\rho \phi) \,\gamma_\mu \psi + i \frac{g}{2} \phi^2 \gamma_\mu (i \gamma_2) \overline{\psi}^T \nonumber \\
\overline{J}_\mu & = & \overline{\psi} \gamma_\mu \, (\gamma_\rho \partial_\rho \phi^{*} )   + i \frac{g}{2} \phi^{*2} {\psi}^T (i \gamma_2) \gamma_\mu \nonumber
\eea

Being conserved, supercharges $J_0, \overline{J}_0$ do not acquire any anomalous dimensions at the critical point and therefore their correlation functions are given by:
$\langle J_0(\tau,\vec{x}) \overline{J}_0(\tau,\vec{x})\rangle  \sim \frac{1}{(x^2 + \tau^2)^2}$.
The supercharge $J_0$ carries electromagnetic charge of three units and transforms in the same way as the microscopic operator $\mathcal{O} \sim c_{\uparrow}(x) c_{\downarrow}(x) c_{\uparrow}(x')$
 where the points $x$ and $x'$ are close to each other. It might be possible to measure the correlation functions of $\mathcal{O}$ in an ARPES experiment similar to
that used to measure cooper-pair correlations \cite{kouzakov2003}.

For completeness, let us briefly mention the results for magnetic phase transitions at the surface of a three dimensional AII TI, as dicussed in Ref.\cite{xu2010}. When the chemical potential is tuned to obtain a Dirac fermion, though the theory is not supersymmetric
(recall that supersymmetry requires equal number of bosonic and fermionic species), the fermions are still strongly coupled to the Ising order parameter and acquire a non-trivial anomalous dimension,
calculable in an $\epsilon$-expansion \cite{zinnjustin, xu2010}.
For a generic chemical potential, one obtains a helical Fermi surface for the Dirac fermions and for weak coupling, the interaction between
the fermions and the critical modes is irrelevant \cite{xu2010} and the magnetic transition remains in the three-dimensional Ising universality class.

\begin{center}
\begin{table*}[ht]
\begin{tabular}{|l|l|l|l|}
\hline
Total spatial & \textbf{Surface Modes at }  & \textbf{Topological Defects at} &  \textbf{Surface-only Criticality} \\
dimension $d \downarrow$ & \textbf{Bulk Criticality} & \textbf{Bulk Criticality} & \\
\hline
$\mathbf{d= 1}$ & TSC Edge modes & & \\
& $\langle \chi(\tau) \chi(0)  \rangle \sim 1/\tau$ & - & - \\
\hline
$\mathbf{d= 2}$ & Non-interacting Majorana modes &  Point defect in TSC: Non-trivial  & \\
& (weak coupling to critical  & fixed point in $\epsilon = 3-d$ expansion: & 1. Ising magnetic ordering  \\
& fluctuations irrelevant) & & of TSC surface:  Tricritical Ising point\\
&  & $\langle \chi(\tau) \chi(0)  \rangle \sim 1/\tau^\eta$ with $\eta \approx 0.32$ &  \\
&  &   & 2. XY-magnetic/superconducting   \\
& & & ordering of  TI surface:\\
& & & central charge $c = 2 \rightarrow 1 $ transition.\\
\hline
$\mathbf{d= 3}$ & Non-interacting Majorana & Line-defect in TSC: & 1. Ising magnetic ordering \\
& modes in TSC and Non-interacting & logarithmic corrections to scaling, & of TSC surface: \\
&Dirac modes/Helical Fermi liquid in  & a `marginal Majorana liquid'. &  $\mathcal{N}=1$ Supersymmetric critical point   \\
& TI (weak coupling to & & 2. Superconducting ordering   \\
&critical fluctuations irrelevant)  & & of TI surface (with chemical potential \\
& & & tuning): $\mathcal{N}=2$ Supersymmetric\\
& & &  critical point.\\
 \hline
\end{tabular}
\caption{Compilation of results discussed in the main text. The three
columns correspond to the three class of problems shown schematically in Fig.\ref{fig:surfacetrans}, Fig.\ref{fig:pointdefect} and Fig.\ref{fig:linedefect}.}
\end{table*}
\end{center}

\underline{(iv) Surface transitions in two dimensional TI:}

In contrast to the above examples, the surface phase transitions in a two dimensional TI coupled to a power-law XY magnetic or superconducting order are of qualitatively different nature.
This is because the one-dimensional edge of a TI can support a stable power-law Berezinskii-Kosterlitz-Thouless  (BKT) phase \cite{bkt} at zero temperature  and therefore, one can potentially find a whole phase (rather than
a fixed point) where the helical fermions and critical bosonic fluctuations are
strongly coupled.  As a first example, consider the case of helical edge modes of TI \cite{wu2006} coupled to critical magnetic fluctuations. The bosonized Hamiltonian is:

\bea
H & = & \int dx_1 \, u K_1\left( \partial_x \theta \right)^2 + \frac{u}{K_1}\left( \partial_x \phi \right)^2 +  K_2 (\partial_x \xi)^2 + \nonumber \\
& &  \frac{1}{K_2}( \partial_x \nu)^2 + r_v cos(\nu) \,+ g \cos(2\phi - \xi ) \label{eq:helical1DTI}
\eea

Here we have bosonized the helical Luttinger liquid with $\psi_{R\uparrow} = e^{i(\theta - \phi)}$, $\psi_{L\downarrow} = e^{i(\theta + \phi)}$ and $[\phi(x),\nabla \theta(x')] = i \pi \delta(x-x') $. $u$ is the
velocity of helical modes, $\xi$ is
the XY magnetic order parameter field while
$e^{i\nu}$ inserts a $2\pi$ vortex in the $\xi$ field and satisfies the commutation relation $ [\xi(x), \nabla \nu(x')] = i \pi \delta(x-x')$. Depending on the magnitude of $K_2$, the vortex term
$r_v$ can be relevant/irrelevant. When $r_v$ is relevant, the power-law  XY-order is lost and the correlations of the $\phi$ field decay exponentially. In this case, the coupling $g$ to the helical
modes is irrelevant and thus the fermionic edge modes are not affected by the critical fluctuations. On the other hand, when $r_v$ is irrelevant, the field $\chi$ is in the BKT phase and
one can always find $K_1,K_2$ such that $g$ is relevant. Relevance of $g$ implies that the fields $2\phi$ and $\xi$ lock together and are described by a single compact variable.
Therefore the central charge of the system jumps from
two to one as $g$ becomes relevant. The analysis for helical Luttinger liquid coupled to
superconducting fluctuations proceeds analogously with the replacement $\phi \rightarrow \theta$ in the interaction term $g$. The
superconducting phase locks to the $2\theta$ and the central charge jumps across the phase transition from two to one, gapping out half the degrees of freedom.

\section{Line/point defects Coupled to Bulk Criticality}

Lastly, we discuss the fate of line and point defects in d=2 and d=3  DIII TSCs respectively, when the system is driven through a magnetic phase transition. Since the defects should preserve time reversal symmetry these cannot be regular vortices. Instead, a time reversal invariant vortex which winds in spin space is required as discussed in \cite{qi2008}. Alternately, a lattice dislocation in a  system with a weak topological index leads to gapless modes in certain cases \cite{RanZhangAV}.
 A general classification of defect modes was presented in Ref.\cite{teo2010}.

\begin{figure}
\begin{centering}
\includegraphics[scale=0.4]{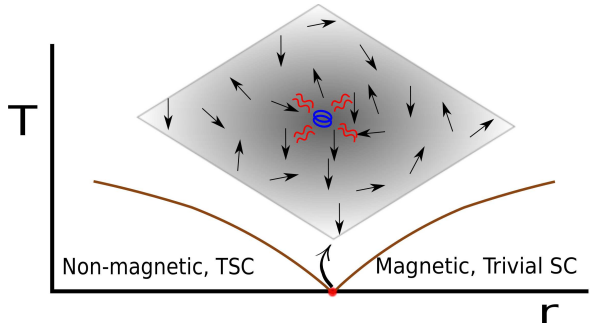}
\par\end{centering}
\caption{ A schematic drawing of a point defect carrying two Majorana modes (blue circles) coupled to critical magnetic fluctuations at the transition between a non-magnetic two-dimensional DIII TSC
and a magnetically ordered ordinary insulator.} \label{fig:pointdefect}
\end{figure}

\begin{figure}
\begin{centering}
\includegraphics[scale=0.4]{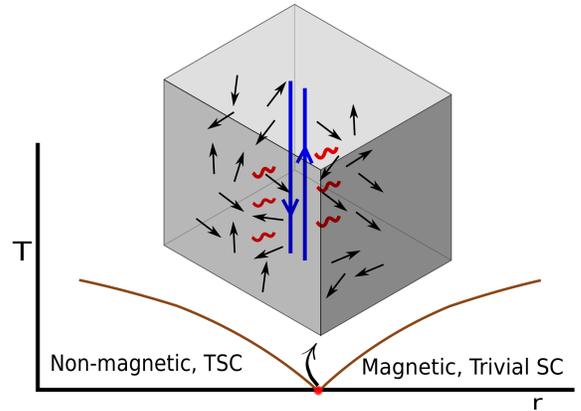}
\par\end{centering}
\caption{ A schematic drawing of a line defect carrying helical Majorana modes (blue lines) coupled to critical magnetic fluctuations at the transition between a non-magnetic three-dimensional DIII TSC
and a magnetically ordered ordinary insulator.} \label{fig:linedefect}
\end{figure}

\underline{(i) Point defects in two-dimensional TSC at criticality:}

 As a first example, consider a point defect in a two-dimensional TSC in class DIII
that hosts two Majorana modes $\chi_1$ and $\chi_2$ which are coupled to Ising critical fluctuations $\phi$ (Fig.\ref{fig:pointdefect}).
The total Euclidean action $S$ of the system near the critical point is given by $S = S_{bulk} + S_{defect} + S_{bulk-defect}$ where

\bea
S_{bulk} & = & \int d^dx\,d\tau \left( \frac{1}{2}(\partial_\mu \phi)^2 + \frac{r}{2} \phi^2 + \frac{u}{4} \phi^4\right) \label{eq:sbulk} \\
S_{defect} & = & \int d\tau \,\, \left( \chi_1 \partial_\tau \chi_1 + \chi_2 \partial_\tau \chi_2 \right) \label{eq:sdefect0d} \\
S_{bulk-defect} & = & i g \, \int d\tau\,\, \chi_1(\tau) \chi_2(\tau) \,\phi(\tau,\vec{x}_\perp = \vec{0})  \label{eq:sbulkdefect0d}
\eea
%
%

Note that in contrast to the case of surface states coupled to critical fluctuations studied above, where one needs to consider the surface critical exponents for the field $\phi$, in this case the relevant exponents
are determined by the conventional bulk criticality. This is because even though the superconducting order parameter vanishes at the location of the topological defect, the defect does not provide any
constraint on the magnetic fluctuations. This is consistent with the conventional treatment of Kondo problem impurity and related problems \cite{wilson1975, andrei1983, hewson2000, sengupta2000, sachdev1999, vojta2007}
where bulk criticality determines the fate of impurity spin.
The differences with the conventional Kondo problem are worth mentioning as well. In particular, the `impurity' in our problem is inherently fermionic,
and satisfies the constraint $\chi_1^2 = \chi^2_2 = 1$ without requiring
any chemical potential. In contrast, in the Kondo problem with, e.g., spin S=1/2 impurity, if one represents the spin as $\vec{S}$ as $\vec{S} = \frac{1}{2} z^{\dagger} \vec{\sigma} z$, where $z$ is a two-component `slave' boson/fermion, one needs to
explicitly  enforce the constraint $z^{\dagger} z = 1$ explicitly using a chemical potential or by other means \cite{abrikosov1965,read1983,coleman1984}. In this sense, our problem has features both of a Kondo impurity
problem and Yukawa-Higgs problem of fermions coupled to bosons \cite{zinnjustin}.
 We will perform a perturbative renormalization group (RG) in the parameter $\epsilon = 3 - D$, the physically relevant case being $\epsilon = 1$. We note that
a potential anisotropy term $\int d\tau\phi^2(\tau, \vec{x}_\perp=\vec{0})$ will also be generated under RG but it is irrelevant since its scaling dimension is unity when $\epsilon = 0$. The RG flow for $g$ is
(Supplementary Information, Sec. B):


\be
\frac{dg}{dl} = \frac{\epsilon g}{2} - \frac{(8-\pi)g^3}{16\pi^3}
\ee

Thus we find a stable infra-red (IR) fixed point at  $g^{*} = \sqrt{8\pi^3 \epsilon/(8-\pi)}$. The anomalous dimension of the Majorana fermion at this fixed point is $\eta_\chi = \frac{g^{*2}}{16\pi^2}
= \frac{\epsilon \pi}{2(8-\pi)}$. Setting $\epsilon = 1$, one finds $\langle \chi(\tau) \chi(0) \rangle \approx 1/\tau^{0.32}$. Therefore, the Majorana fermion behaves like a
zero-dimensional strongly coupled system at criticality, similar to the case of the Majorana edge states of one-dimensional DIII TSC at bulk criticality. A possible physical realization is a dislocation in a two dimensional sheet of the organic superconductor (TMTSF)$_2$PF$_6$ as discussed in the following section.


\underline{(ii) Line defects in three-dimensional TSC at criticality:}

An analogous problem is that of a line defect embedded in a $3+1$-d TSC in class DIII at the quantum criticality between the TSC and an Ising magnet  (Fig.\ref{fig:linedefect}). Away from the criticality,
the line defects in TSC host free helical Majorana modes $\chi_L$ and $\chi_R$ whose Hamiltonian is same as that in Eqn.\ref{eq:helicalmajorana1D}. The bulk action is identical to Eqn.\ref{eq:sbulk} with $d=3$ while

\bea
S_{bulk-defect} & = & 2 i g \int d\tau dx_1\, \chi_L(\tau,x_1) \chi_R(\tau,x_1) \times \nonumber \\
& & \phi(\tau,x_1,\vec{x}_\perp = \vec{0}) \nonumber
\eea

In addition, a potential term $ v \int d\tau dx_1 \phi^2(\tau, x_1, \vec{x}_\perp=\vec{0})$  is also allowed. Unlike the case of point defect embedded in a two dimensional TSC, a simple power-counting shows that in this case,
both the coupling $g$ as well as $v$ are marginal. The RG flow of $g$ and $v$ is given by (see Supplementary
Information, sec. C)

\bea
\frac{dg}{dl} & = & - \frac{3\log(2)}{16\pi^2} g^3 \nonumber \\
\frac{dv}{dl} & = & -\frac{3uv}{64\pi^2} \nonumber
\eea
Because of the marginality of $g$, the correlation function of the Majorana fermions $G(z) = \left< \chi_R(0) \chi_R(z) \right>$ receives logarithmic corrections at criticality:

\bea
G(z) \sim \frac{1}{z(\log(|z|)^{1/6}}
\eea

where $z = x + i \tau$. The correlation function for the left-moving Majorana fermion is simple given by the replacement $z \rightarrow \overline{z}$ in $G$. Therefore, the helical Majorana metal in the
vortex core is a `marginal Majorana liquid' in the quantum critical regime.

A similar problem can be discussed in the context of a dislocation line in a 3dTI, which may carry protected one dimensional modes when a weak index is present \cite{RanZhangAV}. When bulk magnetic ordering sets in via an ising critical point, the fate of coupling the 1dmodes to these critical fluctuations remains to be discussed. Since interactions already modify the electronic properties qualitatively, the answer depends on further details and will be analyzed elsewhere.


%
%

\section{Prospects for Experiments}
To experimentally realize the predictions above, one needs both a symmetry protected topological phase as well as proximity to a quantum critical point where a symmetry
 protecting the topology is spontaneously broken. Topological insulators like Bi$_2$ Se$_3$ doped with magnetic atoms like Fe, Mn, Cr etc. may undergo a magnetic transition
 either just on the surface or in the bulk, on tuning concentration, as recently reported\cite{SurfaceMagneticImpurities1,SurfaceMagneticImpurities2}. Correlated systems such
 as rare earth based half-Heusler materials \cite{half-heusler}, as well as iridium and osmium based oxides may also exhibit a topological insulating phase in proximity to magnetic order\cite{iridates,osmates}.

Some of our most interesting results pertain to topological superconductors with time reversal symmetry (DIII-TSc). A well known physical realization is the B phase of
superfluid He$_3$\cite{He3}. In the solid state context, an interesting proposal that Cu$_x$Bi$_2$Se$_3$ may be a DIII-TSc was recently made\cite{fu2010}, for which there
 is some intriguing, but as yet not conclusive, experimental indications\cite{Ando}. Introducing magnetism into this system remains to be explored.

Among the most tunable materials are rare earth inter-metallics (heavy fermion systems) and organic materials, which show a variety of phases and quantum phase transitions.
 A promising set of candidate materials are UGe$_2$,URhGe and UCoGe \cite{Ucompounds} which show overlap of ferromagnetism and superconductivity in their phase diagrams,
 and which can be tuned by pressure. If the superconductivity, which is likely triplet, is topological in the paramagnetic phase, then this would be a class of candidate
materials\cite{RaghuAV}. Finally we mention the quasi-one dimensional organic superconductor (TMTSF)$_2$PF$_6$ \cite{chaikin}  is believed to be have equal spin triplet
 pairing. The proposed pairing given the highly one dimensional nature of the Fermi surface points to a d=1 topological superconductor in class DIII. A two dimensional sheet of this material will then feature a weak topological index. Interactions with a nearby magnetic phase have also been discussed \cite{chaikin2}. These experimental works contain between them the requisite ingredients to observe the physics discussed in this paper, but more work is of course required to incorporate all desired properties in a single realization.

\section{Conclusions and Perspective}

To summarize, we studied the fate of edge modes and topological defects in TI/TSC in a wide range of phase transitions that do not close the single fermion gap in the bulk.
Our results are summarized in
table I.
One broad and somewhat surprising lesson we learned is that for weak coupling to the critical fluctuations, the edge modes that arise due to topological nature of TI/TSC remain well-defined even at the criticality.
In some cases, such as when the bulk undergoes phase transition in a three dimensional TSC, even
the qualitative nature of the surface states remains unchanged for weak coupling to the critical fluctuations.
While in other cases, such as in the case of Majorana modes at the ends of a one dimensional TSC, the localized modes instead show behavior rather different from free Majorana fermions at the criticality. We also find evidence for
emergent supersymmetry at surface phase transitions in TSC/TI in several cases. Measurable consequences of these results for experiments as well as numerics were discussed and currently available experimental possibilities were reviewed. In future it will be interesting to study the effects of disorder on these conclusions, since this would be relevant to the case where randomly placed impurities induce a magnetic transition. 
To conclude, the study of critical phenomena in topological insulators and superconductors lies at the intersection of
many conceptual and topical themes in condensed matter, such as interacting topological phases, bulk and boundary quantum criticality, and strongly coupled fermionic liquids, while also opening more exotic possibilities such as
the realization of emergent supersymmetry in condensed matter systems.

\underline{Acknowledgements}: We thank Masaki Oshikawa and Mike Zaletel for very helpful discussions and suggestions. Support from NSF DMR- 0645691 is gratefully acknowledged.

\section{Supplementary Information} \label{sec:suppl}

\subsection{A: Duality for 1-d TSC coupled to Critical Ising Chain for a finite chain} \label{sec:openends}

In the main text, we considered the fate of boundary Majorana modes in  a semi-infinite one-dimensional time-reversal invariant TSC at the phase transition
between TSC and an Ising ordered state.  Here we repete a similar analysis for a finite chain that has a pair of Majorana modes at both ends.
The Hamiltonian for the coupled Majorana-Ising system at the bulk criticality is:

\be
H = -\sum_{i=1}^{L} \left( \sigma_i^z \sigma_{i+1}^z + \sigma_i^x \right) - i \sigma_1^z \chi_1 \chi_2 - i \sigma_L^z \chi'_1 \chi'_2
\ee

where we have set the coupling between the Majorana modes and the critical chain to unity (the deviation from unity turns out to be irrelevant at the stable fixed point, via the
same argument as for the semi-infinite chain in the main text). We also ignore the interaction between the Majorana modes at the two ends since it would be suppressed exponentially
in the system size $L$. We define $i \chi_1 \chi_2 = \sz_0$ and $i \chi_1 \chi_2 = \sz_{L+1}$:

\be
H = -\sum_{i=1}^{L} \sigma^{x}_i - \sum_{i=0}^{L} \sigma^{z}_i \sigma^{z}_{i+1}
\ee

$H$ as written above corresponds to transverse field Ising without the transverse field for the first spin $\sigma_0$ and the last spin $\sigma_{L+1}$.
Following \cite{masaki}, let us perform the following duality transformation:

\bea
\sigma^{x}_0 & = & \mu^{z}_0 \\
\sigma^{x}_i & = & \mu^{z}_{i-1} \mu^{z}_{i} \hspace{0.5cm} \,\forall\,\, \,L+1 \geq i \geq 1 \\
\sz_i \sz_{i+1} & = & \mx_i \hspace{0.5cm} \forall\,\, L \geq i \geq 0 \\
\sz_{L+1} & = & \mx_{L+1}
\eea

Hamiltonian $H$ in terms of new variables $\mu$ is given by the standard transverse field Ising model:

\be
H = -\sum_{i=0}^{L} \left( \mx_i + \mz_i \mz_{i+1} \right)
\ee

The unequal time correlation function for the Majorana modes at the two ends are given by $\langle \chi_\alpha(0) \chi_\alpha(\tau) \rangle = \langle \sx_0(0) \sx_0(\tau) \rangle =
\langle \mz_0 (0) \mz_0(\tau) \rangle \sim 1/\tau$ and  $\langle \chi'_\alpha(0) \chi'_\alpha(\tau) \rangle = \langle \sx_{L+1}(0) \sx_{L+1}(\tau) \rangle =
\langle \mz_L (0) \mz_L(\tau) \rangle \sim 1/\tau$ for both $\alpha = 1,2$.
\subsection{B: Renormalization for point defect in two dimensional TSC coupled to critical fluctuations} \label{sec:rgpoint}

Since the bulk is unaffected by the edge modes, we first imagine that we have already performed renormalization for the bulk degrees of the freedom and are sitting at the bulk criticality.
The RG equations for $S_{bulk}$ to $O(\epsilon)$ are \cite{Goldenfeld}

\bea
\frac{du}{dl} & = & \epsilon u - \frac{9 u^2}{8 \pi^2} \\
\frac{dr}{dl} & = & 2r + \frac{3u\Lambda^2}{8\pi^2}-\frac{3ur}{8\pi^2}
\eea

where $\Lambda$ is a UV cut-off. This implies that there is a stable IR fixed point (with one relevant direction) at $(r^{*}, u^{*}) = \left(\frac{-\epsilon \Lambda^2}{6}, \frac{8 \pi^2 \epsilon}{9} \right)$.

To perform Wilsonian RG for $S_{edge} + S_{bulk-edge}$, we separate the degrees of freedom into `slow' and `fast' i.e. we write
\bea
\chi(x_0) & = & \int_{|k_0| < \Lambda/s} dk_0 \,\,  \chi(k_0) e^{i k_0 x_0} + \int_{\Lambda< |k_0| < \Lambda/s} dk_0\,\, \chi(k_0) e^{i k_0 x_0}  \nonumber \\
& = & \chi_{<}(x_0) + \chi_{>}(x_0)
\eea

Similarly, we define,

\bea
\phi(x_0, \vec{r}_{\perp}=0) & = & \int_{|k_0| < \Lambda/s} dk_0\,d^{d-1}k_{\perp} \,\,  \phi(k_0,\vec{k}_\perp) e^{i k_0 x_0} + \nonumber \\
& & \int_{\Lambda< |k_0| < \Lambda/s} dk_0\,d^{d-1}k_{\perp}\,\, \phi(k_0,\vec{k}_\perp) e^{i k_0 x_0} \nonumber \\
& = & \phi_{<}(\vec{r}) + \phi_{>}(\vec{r})
\eea

where $\vec{r} = \left(x_0, \vec{r}_\perp \right)$. Note, in particular, that the perpendicular momenta $\vec{k}_\perp$ are being integrated over the whole range in both $\phi_{<}$ and $\phi_{>}$. This is legitimate because $|\vec{k}_\perp| \approx 0$
do not generate any singularities for the edge action \textit{as long as the modes near $k_0 \sim 0$ are never integrated over}, which is indeed true owing to the above separation.

The action $S$ in terms of new variables $\phi_{>},\phi_{<},\chi_{>},\chi_{<}$ is

\bea
S = S^{<}_{0} + S^{>}_{0} + S_1
\eea

where $S_{0}$ denotes the Gaussian part of the full action and which, therefore, separates into the slow and fast modes without any cross-terms. $S_1$, the interaction part includes the
$\phi^4$ term as well as $S_{bulk-edge}$. Since the $\phi^4$ part is unaffected by the edge modes, we will only focus on the $g$ term as far as $S_1$ is concerned. The full partition function is given by

\bea
Z & = & \int D\phi_<\, D\chi_{1<} \, D\chi_{2<}\, e^{-S^{<}_0} \int D\phi_> \, D\chi_{1>} \, D\chi_{2>}\, e^{-S^{>}_0} e^{-S_1} \\
& = & Z^{>}_0 \int D\phi_<\, D\chi_{1<} \, D\chi_{2<} e^{-S^{<}_0} \left<  e^{-S_1}\right>_>
\eea

where $Z^{>}_0 =  \int D\phi_> \, D\chi_{1>} \, D\chi_{2>}\, e^{-S^{>}_0}$ and

\be
\left<  e^{-S_1}\right>_> = \frac{1}{Z^{>}_0}\int D\phi_> \, D\chi_{1>} \, D\chi_{2>}\, e^{-S^{>}_0} e^{-S_1}
\ee

The integration over the fast modes $\phi_>, \chi_{1>}, \chi_{2>}$ would modify the action for the slow modes  $\phi_<, \chi_{1<}, \chi_{2<}$, thus yielding the RG equations so desired.
Since we suspect a fixed point at $g^{*} = O(\sqrt{\epsilon})$, a cumulant expansion in powers of $g$ is a legitimate option. Therefore, upto $O(g^3)$, $\left<  e^{-S_1}\right>_>$ may be written as

\be
\left<  e^{-S_1}\right>_> = e^{-\left( \left< S_1\right>_> - \frac{1}{2}\left[ \left< S^2_1\right>_> - \left< S_1\right>_>^2 \right] + \frac{1}{6}\left[
\left< S^3_1\right>_> - 3\left< S^2_1\right>_> \left< S_1\right>_> + 2 \left< S_1\right>_>^3 \right]\right)}
\ee

We next write down the contribution from each of
 the three terms in the cumulant expression to the renormalization of the $S_{bulk-defect}$ as well as the anomalous dimension of Majorana fermions.

\vspace{0.5cm}

$\boxed{O(g):\,\left<S_1\right>_>}$

\bea
\left<S_1\right>_> = i g \, \int dx_0\,\, \chi_{1<}(x_0) \chi_{2<}(x_0) \, \phi_<(x_0,\vec{x}_\perp = \vec{0})
\eea

Consider the elementary coarse-graining transformation $x'_0 = x_0/s$. Under this transformation, $\chi_{1<}(x_0) = \xi_\chi \chi_1(x'_0,)$, $\chi_{2<}(x_0) = \xi_\chi \chi_2(x'_0,)$ and
$\phi_<(x_0,\vec{x}_\perp=\vec{0}) = \xi_\phi \phi(x'_0,\vec{x}_\perp=\vec{0})$. $\xi_\chi$ and $\xi_\phi$ are related to the scaling dimensions $\Delta_\chi, \Delta_\phi$ of the fields $\xi$ and $\chi$
for $d = 4-\epsilon$ via $\xi_{\chi} = s^{-\Delta_\chi} = s^{-\eta_\chi/2}$ and $\xi_{\phi} = s^{-\Delta_\phi} = s^{-(d-1+\eta_\phi)/2}$. Re-expressing $\left<S_1\right>_>$ in terms of rescaled variables,

\bea
\left<S_1\right>_> = i g s^{\epsilon/2 -\eta_\chi} \, \int dx'_0\,dx'_1\,\, \widetilde{\chi ^T}(x'_0,x'_1) \gamma_S \chi(x'_0,x'_1) \, \phi(x'_0,x'_1,\vec{x}_\perp = \vec{0})
\eea

where we have dropped the contribution from $\eta_\phi$ since it is of order O($\epsilon^2$) (we will soon see that $\eta_\chi = O(\epsilon)$). Therefore, at this order, $g(s) = g s^{\epsilon/2-\eta_\chi}$ where $\epsilon = 3-d$.

\vspace{0.5cm}

$\boxed{O(g^2):\,- \frac{1}{2}\left[ \left< S^2_1\right>_> - \left< S_1\right>_>^2 \right]}$

These terms renormalize the propagator of the edge $\chi$ fermions
leading to an anomalous dimension for the fermions as shown in diagram. Denoting this contribution by $I_\chi$,

\bea
I_\chi & = &\frac{1}{2} \times \frac{g^2}{(2\pi)^4} \int_{0<|k_\perp|<\Lambda} \int_{\Lambda/s<|k_0|<\Lambda} \frac{d \,k_0 \,\,d^d \,k_\perp}{(k_0 + p_0)(k^2_0 + k^2_\perp + r^{*})} \\
& = & \frac{g^2}{16 \pi^2} p_0\, \log(s)
\eea

Hence $\eta_\chi = \frac{g^2}{16 \pi^2}$.
\vspace{0.5cm}

$\boxed{O(g^3):\,\frac{1}{6}\left[
\left< S^3_1\right>_> - 3\left< S^2_1\right>_> \left< S_1\right>_> + 2 \left< S_1\right>_>^3 \right]}$


Only the term $\left<S^3_1\right>_>$ contributes to the renormalization of the interaction $g$ at this order as depicted
in Fig.\ref{feyn_g}. Let us call these contribution $\Delta S_g$.

\bea
\Delta S_g =  \times (ig)^3 \, \int dx_0\,\, \chi_{1<}(x_0) \chi_{2<}(x_0) \, \phi_<(x_0,\vec{x}_\perp = \vec{0}) \times I
\eea

where

\bea
I & = & \frac{1}{(2\pi)^{4}} \int_{\Lambda/s<|p|<\Lambda} dp_0\,\int_{0<|p_\perp|<\Lambda}d^{3}p_\perp \,\,\frac{1}{p_0(p_0^2+p_\perp^2+r^{*})} \\
& \approx & \frac{1}{2\pi^3}\times(1-\pi/4) \times \frac{s-1}{s}
\eea

Using this, one finds

\bea
g(s) = s^{\epsilon/2 - \frac{g^2}{16 \pi^2}}\left(g-\left(\frac{1-\pi/4}{2\pi^3}\right)\left(\frac{s-1}{s}\right)g^3\right)
\eea

Writing $s = e^{dl}$, the RG flow for g becomes
\be
\frac{dg}{dl} = \frac{\epsilon g}{2} - \frac{(8-\pi)g^3}{16\pi^3}
\ee

Thus, we find a stable IR fixed point at $g^{*} = \sqrt{8\pi^3 \epsilon/(8-\pi)}$.

The anomalous dimension of the Majorana fermion is given by

\bea
\eta_\chi & = & \frac{g^{*2}}{16\pi^2}\\
& = & \frac{\epsilon \pi}{2(8-\pi)}
\eea

Setting $\epsilon = 1$, one finds $\eta_\chi = \frac{\pi}{2(8-\pi)} \approx 0.32$.

\subsection{C: Renormalization for line defect in three dimensional TSC coupled to critical fluctuations} \label{rgline}

The total Euclidean action of the system near the critical point is given by:

\be
S = S_{bulk} + S_{defect} + S_{bulk-defect}
\ee

where $S_{bilk}$ is same as in the case of point defect with $d=3$ and

\bea
& & S_{defect}  \nonumber \\
&= & \int dx_0\,dx_1 \,\left( \chi_R \partial_\tau \chi_R + \chi_L \partial_\tau \chi_L + i \chi_R \partial_x \chi_R - i \chi_L \partial_x \chi_L \right) \nonumber \\
& = & \int dx_0\,dx_1 \,\, \widetilde{\chi ^T} \gamma_\mu \partial_\mu \chi
\eea

where the matrices $\gamma_0 = \tau_x, \gamma_1 = \tau_y$ act on the vector $\chi = [\chi_R \, \chi_L]^T$ and  $\widetilde{\chi ^T} = \chi ^T \gamma_0$. The coupling $S_{bulk-defect}$ between the critical bulk
and boundary CFT is given by

\bea
& & S_{bulk-defect} \nonumber \\
& = & 2 i g \, \int dx_0\,dx_1\,\, \chi_L(x_0,x_1) \chi_R(x_0,x_1) \,\phi(x_0,x_1,\vec{x}_\perp = \vec{0}) \nonumber \\
& = & i g \, \int dx_0\,dx_1\,\, \widetilde{\chi ^T}(x_0,x_1) \gamma_S \chi(x_0,x_1)\, \phi(x_0,x_1,\vec{x}_\perp = \vec{0}) \nonumber
\eea

where $\gamma_S = \tau^z = -i \gamma_0 \gamma_1$. In addition, one generates a potential anisotropy term $v \int dx_0 dx_1 \phi^2(x_0,x_1, x_\perp = 0)$ localized near the defect.

Again the bulk criticality is unaffected by the boundary.
Similar to the case of point defect, we write

\bea
\chi(\vec{r}) & = & \int_{|\vec{k}| < \Lambda/s} d^2k \,\,  \chi(\vec{k}) e^{i \vec{k}.\vec{r}} + \int_{\Lambda< |\vec{k}| < \Lambda/s} d^2k\,\, \chi(\vec{k}) e^{i \vec{k}.\vec{r}} \nonumber \\
& = & \chi_{<}(\vec{r}) + \chi_{>}(\vec{r})
\eea

where $\vec{r} = \left(x_0, x_1\right)$ and

\bea
\phi(\vec{r}, \vec{r}_{\perp}=0) & = & \int_{|\vec{k}| < \Lambda/s} d^2k\,d^{d-2}k_{\perp} \,\,  \phi(\vec{k},\vec{k}_\perp) e^{i \vec{k}.\vec{r}} \nonumber \\
& & + \int_{\Lambda< |\vec{k}| < \Lambda/s} d^2k\,d^{d-2}k_{\perp}\,\, \phi(\vec{k},\vec{k}_\perp) e^{i \vec{k}.\vec{r}} \nonumber \\
& = & \phi_{<}(\vec{r}) + \phi_{>}(\vec{r})
\eea

The full partition function is given by

\bea
Z & = & \int D\phi_< D\chi_< D{\widetilde{\chi}_< ^T} e^{-S^{<}_0} \int D\phi_> D\chi_> D{\widetilde{\chi}_> ^T} e^{-S^{>}_0} e^{-S_1} \\
& = & Z^{>}_0 \int D\phi_< D\chi_< D{\widetilde{\chi}_< ^T} e^{-S^{<}_0} \left<  e^{-S_1}\right>_>
\eea

where $Z^{>}_0 =  \int D\phi_> D\chi_> D{\widetilde{\chi}_> ^T} e^{-S^{>}_0}$ and

\be
\left<  e^{-S_1}\right>_> = \frac{1}{Z^{>}_0}\int D\phi_> D\chi_> D{\widetilde{\chi}_> ^T} e^{-S^{>}_0} e^{-S_1} \nonumber
\ee

Upto $O(g^3)$, $\left<  e^{-S_1}\right>_>$ may be written as

\be
\left<  e^{-S_1}\right>_> = e^{-\left( \left< S_1\right>_> - \frac{1}{2}\left[ \left< S^2_1\right>_> - \left< S_1\right>_>^2 \right] + \frac{1}{6}\left[
\left< S^3_1\right>_> - 3\left< S^2_1\right>_> \left< S_1\right>_> + 2 \left< S_1\right>_>^3 \right]\right)}
\ee

We next write down the contributions from each of
 the three terms in the cumulant expression.

\vspace{0.5cm}

$\boxed{O(g):\,\left<S_1\right>_>}$

\bea
\left<S_1\right>_> = i g \, \int dx_0\,dx_1\,\, \widetilde{\chi ^T}_<(x_0,x_1) \gamma_S \chi_<(x_0,x_1) \, \phi_<(x_0,x_1,\vec{x}_\perp = \vec{0}) \nonumber
\eea

Consider the elementary coarse-graining transformation $x'_0 = x_0/s, \,x'_1=x_1/s$. Under this transformation, $\chi_<(x_0,x_1) = \xi_\chi \chi(x'_0,x'_1)$ and
$\phi_<(x_0,x_1,\vec{x}_\perp=\vec{0}) = \xi_\phi \phi(x'_0,x'_1,\vec{x}_\perp=\vec{0})$. $\xi_\chi$ and $\xi_\phi$ are related to the scaling dimensions $\Delta_\chi, \Delta_\phi$ of the fields $\xi$ and $\chi$
at the free fixed point via $\xi_{\chi} = s^{-\Delta_\chi} = s^{-(1/2 + \eta_\chi/2)}$ and $\xi_{\phi} = s^{-\Delta_\phi} = s^{-(d-1+\eta_\phi)/2}$. Re-expressing $\left<S_1\right>_>$ in terms of rescaled variables,

\bea
\left<S_1\right>_> = i g s^{\epsilon/2-\eta_\chi} \, \int dx'_0\,dx'_1\,\, \widetilde{\chi ^T}(x'_0,x'_1) \gamma_S \chi(x'_0,x'_1) \, \phi(x'_0,x'_1,\vec{x}_\perp = \vec{0}) \nonumber
\eea

Therefore, at this order, $g(s) = g s^{\epsilon/2-\eta_\chi}$ where $\epsilon = 3-d$.

\vspace{0.5cm}

$\boxed{O(g^2, uv):\,- \frac{1}{2}\left[ \left< S^2_1\right>_> - \left< S_1\right>_>^2 \right]}$

These terms renormalize the propagator of the edge $\chi$ fermions
leading to an anomalous dimension for the fermions as shown in diagram. Denoting this contribution by $I_\chi$,

\bea
I_\chi & = & \frac{g^2}{(2\pi)^4} \int_{0<|k_\perp|<\Lambda} \int_{\Lambda/s<|k_\parallel|<\Lambda} \frac{d^2 \,k_\parallel \,\,d^2 \,k_\perp}{(k_\mu \gamma_\mu + p_\mu \gamma_\mu)(k^2 + k^2_\perp )} \nonumber \\
& = & \frac{g^2}{16 \pi^2} p_\mu \gamma_\mu \log(s) \log(2)
\eea

Hence $\eta_\chi = \frac{g^2}{16 \pi^2}$.

One also generates renormalization of $v$:
\bea
\Delta S & = & - \frac{6uv}{8} \,\,\phi^2_<(x_0,x_1,x_\perp=0) \times \nonumber \\
& & \frac{1}{(2\pi)^4}\int_{0<|k_\perp|<\Lambda} \int_{p<|k_\parallel|<\Lambda} \frac{d \,k_\parallel \,\,d^d \,k_\perp}{(k^2 + k^2_\perp )^2} \nonumber
\eea

This leads to

\be
\frac{dv}{dl} = - \frac{3}{64 \pi^2} uv
\ee
\vspace{0.5cm}

$\boxed{O(g^3):\,\frac{1}{6}\left[
\left< S^3_1\right>_> - 3\left< S^2_1\right>_> \left< S_1\right>_> + 2 \left< S_1\right>_>^3 \right]}$

\begin{figure}
\begin{centering}
\includegraphics[scale=1.0]{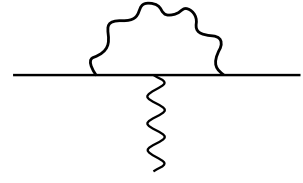}
\par\end{centering}
\caption{The Feynman diagram that contributes to the renormalization of vertex $g$ at $O(g^3)$ for the problem of a line defect coupled to critical fluctuations. The wavy line represents the propagator $\Pi_{\phi}(\vec{p},\vec{p}_\perp) = \frac{1}{p^2 + p^2_\perp + r}$
for the bulk $\phi$ field while the straight line corresponds to the propagator $\Pi_\chi = \frac{1}{i p_\mu \gamma_\mu}$ for the edge fermions. } \label{feyn_g}
\end{figure}

$\left<S^3_1\right>_>$ contributes to the renormalization of the interaction $g$ at this order as depicted
in Fig.\ref{feyn_g}:

\bea
\Delta S_g =  (ig)^3 \, \int dx_0\,dx_1\,\, \widetilde{\chi ^T}_<(x_0,x_1) \gamma_S \chi_<(x_0,x_1) \, \phi_<(x_0,x_1,\vec{x}_\perp = \vec{0}) \times I \nonumber
\eea

where

\bea
I & = & \frac{1}{(2\pi)^{4}} \int_{\Lambda/s<|p|<\Lambda} d^2p\,\int_{0<|p_\perp|<\Lambda}d^{d-1}p_\perp \,\,\frac{1}{p^2(p^2+p_\perp^2+r^{*})} \nonumber \\
& \approx & \frac{\log(2)}{8\pi^2} \times \frac{s-1}{s}
\eea

Collecting everything,

\bea
g(s) = s^{\epsilon/2 -  \frac{g^2}{16 \pi^2}}\left(g-\frac{\log(2)}{8\pi^2}\frac{s-1}{s}g^3\right)
\eea

Writing $s = e^{dl}$, the RG flow for g becomes
\be
\frac{dg}{dl} = \frac{\epsilon g}{2} - \frac{3\log(2)}{16\pi^2} g^3
\ee

For the problem of line defect in a three dimensional TSC, $\epsilon = 0$, hence

\be
\frac{dg}{dl} = - \frac{3\log(2)}{16\pi^2} g^3
\ee

To the leading logarithmic correction, the renormalization group equation for the correlation function $G(z,g) = \left< \chi_R(0) \chi_R(z) \right>$ is given by

\be
\frac{\partial G}{\partial \log(z)} + \beta(g) \frac{\partial G}{\partial g} + (1+Bg^2)G = 0
\ee

where $\beta(g) = \frac{d\, g(|z|)}{d\, \log(|z|)} = - A g^3$ with $B = \frac{\log(2)}{16\pi^2}$ and $A = \frac{3\log(2)}{16\pi^2}$. Solving the above differential equation, one finds

\bea
G(z) & = & \frac{1}{z(\log(|z|)^{\frac{B}{2A}}} \\
& = & \frac{1}{z(\log(|z|)^{1/6}}
\eea

For the left-moving fermions, one simply replaces $z \rightarrow \overline{z}$.

\begin{figure}
\begin{centering}
\includegraphics[scale=1.0]{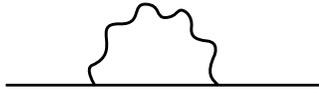}
\par\end{centering}
\caption{The Feynman diagram that contributes to the self-energy of Majorana fermions for the problem of a line defect coupled to critical fluctuations.} \label{self_e}
\end{figure}

\end{document}